\documentclass[aps,prl,reprint,amsfonts,groupedaddress,superscriptaddress]{revtex4-2}
\usepackage[colorlinks,urlcolor=blue,anchorcolor=blue,linkcolor=blue,citecolor=blue,breaklinks=true]{hyperref}
\usepackage{color}
\usepackage{graphicx}

\begin{document}
\title[Magnetically tuned topological phase in graphene nanoribbon heterojunctions]{Magnetically tuned topological phase in graphene nanoribbon heterojunctions}


\author{Wei-Jian Li}
\affiliation{National Laboratory of Solid State Microstructures and Department of Physics, Nanjing University, Nanjing, 210093, Jiangsu, China.}
\author{Da-Fei Sun}
\affiliation{National Laboratory of Solid State Microstructures and Department of Physics, Nanjing University, Nanjing, 210093, Jiangsu, China.}
\author{Sheng Ju}
\email{jusheng@suda.edu.cn}
\affiliation{School of Optical and Electronic Information, Suzhou City University, Suzhou 215104, China}
\affiliation{Jiangsu Key Laboratory and Suzhou Key Laboratory of Biophotonics, Suzhou City University, Suzhou 215104, China}
\author{Ai-Lei He}
\email{heailei@yzu.edu.cn}
\affiliation{College of Physics Science and Technology, Yangzhou University, Yangzhou 225002, China}
\author{Yuan Zhou}
\email{zhouyuan@nju.edu.cn}
\affiliation{National Laboratory of Solid State Microstructures and Department of Physics, Nanjing University, Nanjing, 210093, Jiangsu, China.}
\affiliation{Collaborative Innovation Center of Advanced Microstructures, Nanjing University, Nanjing, 210093, Jiangsu, China.}

\date{\today}
\begin{abstract}
The interplay between topology and magnetism often triggers the exotic quantum phases. Here, we report an accessible scheme to engineer the robust $\mathbb{Z}_{2}$ topology by intrinsic magnetism, originating from the zigzag segment connecting two armchair segments with different width, in one-dimensional graphene nanoribbon heterojunctions. Our first-principle and model simulations reveal that the emergent spin polarization substantially modifies the dimerization between junction states, forming the special SSH mechanism depending on the magnetic configurations. Interestingly, the topological phase in magnetic state is only determined by the width of the narrow armchair segment, in sharp contrast with that in the normal state. In addition, the emergent magnetism increases the bulk energy band gap by an order of magnitude than that in the nonmagnetic state. We also discuss the $\mathbb{Z}$ topology of the junction states and the termination-dependent of topological end states. Our results bring new way to tune the topology in graphene nanoribbon heterostructure, providing a new platform for future one-dimensional topological devices and molecular-scale spintronics.
\end{abstract}
\maketitle
\emph{Introduction}---Topological band theory has been instrumental in predicting and explaining exotic quantum states, such as the quantum spin hall state in topological insulators~\cite{Hasan-RMP2010,Qi-RMP2011}. Topological classification relies on topological order protected by a specified symmetry. Symmetry protected topological (SPT) phase transitions were observed in artificial semiconducting systems such as two-dimensional quantum wells of HgTe~\cite{Konig-Science2007}. Since the discovery of topological quantum phenomenon, progresses on the topology have been mostly focused on two-dimensional and three-dimensional systems for years. In fact, topological quantum phase can also been realized in one-dimensional ($1$D) systems. It was first reported in $1$D polyacetylene described by Su-Schrieffer-Heeger (SSH) model~\cite{Su-PRL1979,Su-PRB1980}, and further discovered in $1$D acenemotif-derived polymers via tailoring their $\pi$-conjugation~\cite{Cirera-NN2020}. Recently, interest has surged in the topology in $1$D graphene nanoribbon (GNR) heterostructures through the on-surface synthesis~\cite{Cai-Nature2010,Ruffieux-Nature2016,Sun-AM2020}. Those $1$D heterojunctions exhibit rich topological physics distinct from the higher-dimensional counterparts~\cite{Cao-PRL2017,Rizzo-Nature2018,Groning-Nature2018,Lee-NL2018,Lin-NL2021,Li-NC2021,Jiang-NL2021,Zhao-PRL2021}. In particular, the zero-dimensional ($0$D) topological end states or junction states, inherited from the $1$D bulk, is dictated by the terminating unit cell~\cite{Cao-PRL2017}, in analogy to the $0$D corner states in the two-dimensional second-order topological insulators~\cite{Benalcazar-Science2017,Sheng-PRL2019,Qian-PRB2021,Wang-arXiv2024}. SPT phase in GNRs, classified by the inversion/mirror symmetry or chiral symmetry(the A/B sublattice symmetry), can be characterized by the $\mathbb{Z}_2$ or $\mathbb{Z}$ invariants~\cite{Cao-PRL2017,Jiang-NL2021}. The $1$D topological phases in GNRs can be further tuned by applying transverse electric field or strain~\cite{Lawrence-ACSN2020,Zhao-PRL2021}. However, those proposed $1$D GNRs feature the relatively narrow bandgap, especially for large-scale ribbons, raising challenges in stabilizing and efficiently engineering $1$D topological phases. How to improve the stability of $1$D topological phases and how to efficiently engineer $1$D topology are still the open questions.

On the other hand, the interplay between topology and magnetism is a key focus in recent condensed matter physics. The coexistence of magnetic order and non-trivial topology often gives rise to fruitful exotic quantum states, including quantum anomalous Hall effect~\cite{PhysRevLett.61.2015,Chang-RMP2023}, axion insulator states~\cite{PhysRevB.78.195424,Mogi-NM2017} and Majorana fermions~\cite{Read-PRB2000,Kitaev-PU2001,PhysRevLett.100.096407,PhysRevLett.102.187001,Heimes-PRB2014,prada2020andreev}. Inducing magnetism into the topologically non-trivial GNRs may allow to design $1$D topological devices, spin-selective devices, symmetry-protected spin-correlated end states or Majorana fermions in close proximity to a superconductor, crucial for developing molecular-scale spintronics\cite{wang2021graphene,kitaev2001unpaired,PhysRevB.97.041414,PhysRevB.102.165147}.  Armchair (edge) GNRs (AGNRs) usually connect to topology depending on the ribbon width and terminations of the unit cell~\cite{Son-PRL2006,Cao-PRL2017}, while the zigzag (edge) GNRs (ZGNRs) often generate magnetism due to the existence of the half-filled partially flat band~\cite{Fujita-JPSJ1996,Son-PRL2006,Magda-Nature2014,Chen-NL2017}. The geometry of GNR heterojuctions (GNRHs) naturally contain both types of edges, providing a new platform to investigate the magnetically tuned topology. However, the influence of induced magnetism on $1$D topology in GNRHs remains unclear.

In this paper, we propose a new scheme to realize the magnetically tuned topology in $1$D GNRHs. The zigzag, and armchair segment offer the magnetism, and initial topology, respectively. First-principle and model simulations show that the number and chirality of $\mathbb{Z}$ topological junction states are insensitive to the emergent magnetism. However, the bulk $\mathbb{Z}_{2}$ topology depends solely on the width of narrow armchair segment, in sharp contrast with that in the nonmagnetic state. Moreover, the energy gap in the magnetic configuration increases an order of magnitude than that in the normal state. We identify that the spin-polarized edge modes along the zigzag edge dramatically change the positions of the Wannier centers(WC), providing a magnetic configuration dependent SSH mechanism. Furthermore, we show that the topological end states are highly sensitive to the termination geometry, reflecting its $1$D topological nature. 

\begin{figure}[!htb]
  \centering
  \includegraphics[width=\linewidth]{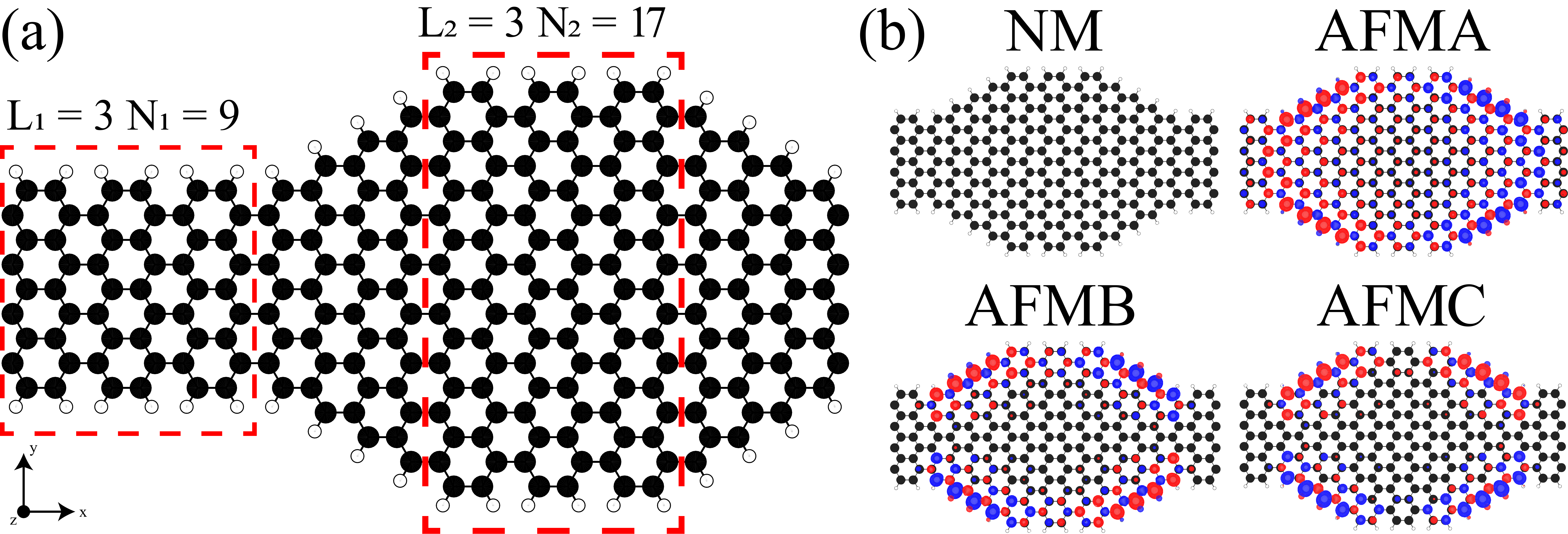}
  \caption{(a) Typical unit cell of the $L_{1}$-$L_{2}$-$N_{1}$-$N_{2}$-GNRH. Here, $3$-$3$-$9$-$17$-GNRH is schematically shown. Black circles and white circles represent carbon atom and passivated hydrogen atoms, respectively. (b) Potential magnetic configurations. Red and blue circles stand for spin up and spin down, respectively. The size of colored circles denote the relative magnitude of spin polarization.}\label{fig1}
\end{figure}

\emph{Model and electronic structure}---The simplest idea to construct the magnetically tuned GNRHs is connecting two AGNR segments with different width by the zigzag segment, while preserving the inversion symmetry. The designed unit cell of the GNRH is denoted as $L_1$-$L_2$-$N_1$-$N_2$-GNRH, where $L_1$, $L_2$, $N_1$, $N_2$ represents the length and width of the narrow and wider AGNR segments as schematically shown in Fig.~\ref{fig1}. The length of ZGNR segment is therefore $\frac{ N_{2}-N_{1}}{2}$. Observable magnetization along the zigzag edge requires at least $3$ carbon atoms, \emph{i.e.}, $N_2-N_1 > 4$. 
We explore the magnetic and topological properties by the first-principle and model simulations (Supplemental Materials (SM) for details).

For simplicity, we use $2$-$3$-$9$-$17$-GNRH as a minimal model. Magnetization arises at the zigzag edges due to local sublattice imbalance~\cite{Fujita-JPSJ1996,Magda-Nature2014}. At half-filling, the total net spin polarization is zero according to the Lieb's theorem on bipartite lattice~\cite{Lieb-PRL1989}, yielding the non-magnetic or antiferromagnetic ground states.
The four possible magnetic configurations preserves the time reversal symmetry, one non-magnetic(NM) and three different antiferromagnetic states, labeled as AFMA, AFMB and AFMC (Fig.~\ref{fig1}(b)). While the AFMA configuration breaks both the inversion and mirror symmetries, precluding the emergence of an SPT phase, the other three preserve the inversion symmetry or mirror symmetry (or both). The total energies of these configurations is less than $0.5$ meV per carbon atom, making those magnetic configurations highly tunable by external fields (Tab. S1 in SM). The bandgap of the NM configuration is about $17$ meV. In contrast, it enhances about an order of magnitude in all magnetic configurations. This means the emergent magnetism in GNRH can significantly improve the stability of potential topology. We notice that the bandstructure near Fermi level in NM, AFMB, and AFMC configuration exhibits the particle-hole-like symmetry, suggesting the possibility of SPT phase in those configurations (Fig. S1 in SM).

\emph{Topological junction states}---Joining two AGNRs belonging to different topological classes creates topological junction states. The number and the chirality of the junction states can be characterized by the topological invariants of the AGNR components. In our designed heterojunctions, the chiral symmetry ($A$ and $B$ sublattice symmetry) inherited from the original graphene honeycomb lattice is preserved, we therefore adopt the $\Delta\mathbb{Z}=\lfloor\frac{N_2}{3}\rfloor - \lfloor\frac{N_1}{3}\rfloor$ to characterize the topological index of the junction states(SM for details). Since we consider $N_{2}-N_{1}\geq 4$ cases, $\Delta\mathbb{Z} \neq 0$, \emph{i.e.}, the topological junction states always exist. We notice that the topological index of junction state is independent the width of GNRs. For example, there are $4$ carbon atoms pairs not connected by $\sigma$-bonds in $8$-$8$-$9$-$17$-GNRH, yielding the topological index for junction state $\Delta\mathbb{Z}=2$. This means that there are two topological junction states located at $A$ sublattice due to the positive value. In contrast, $\Delta\mathbb{Z}=1$ for $8$-$8$-$9$-$13$-GNRH, suggesting one topological junction state located at $A$ sublattice.

\begin{figure}[!htb]
  \centering
  \includegraphics[width=\linewidth]{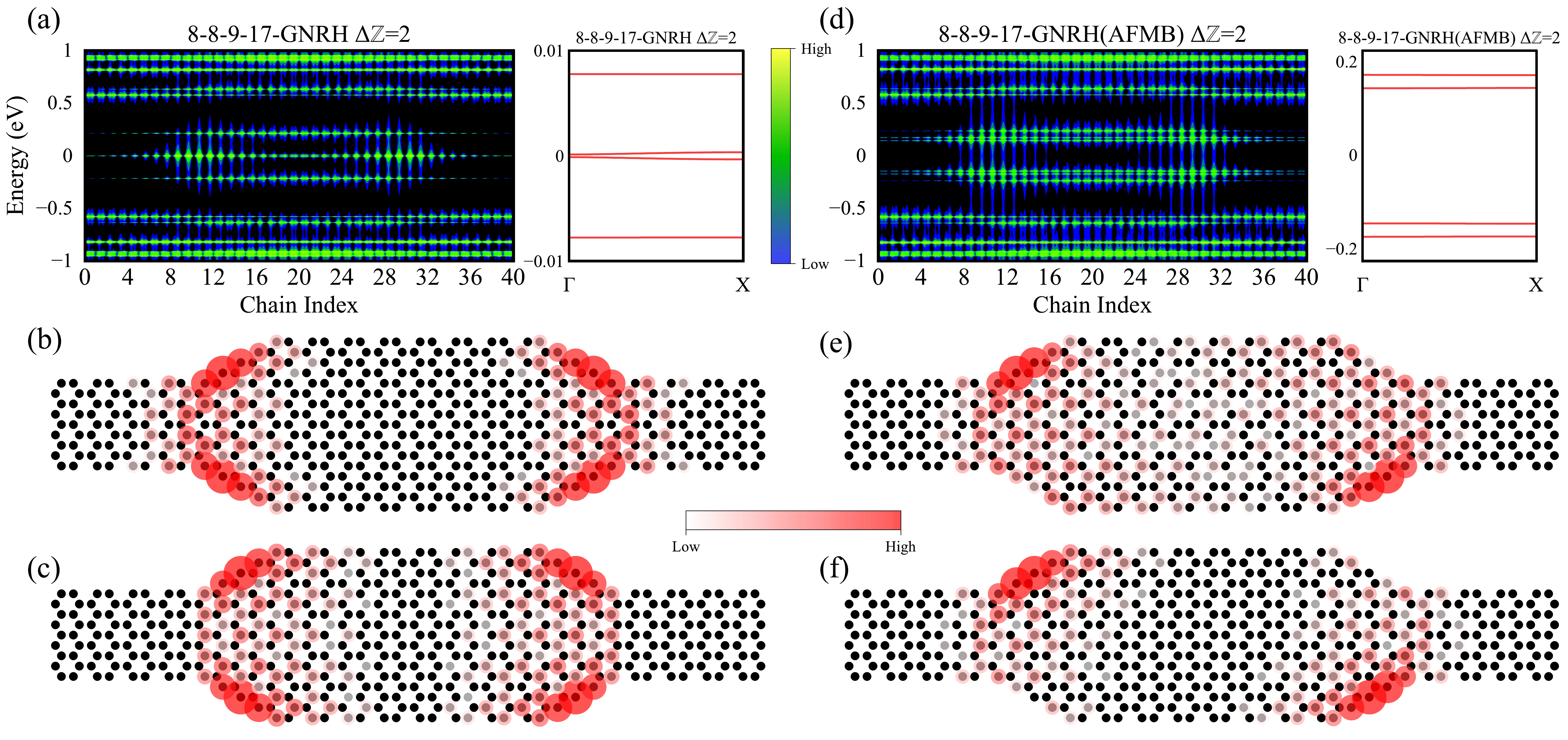}
  \caption{Energy spectrum of $8$-$8$-$9$-$17$-GNRH. Left, and right panels are for NM, and AFMB configurations, respectively. Top panels are for the LDOS and energy spectrum maps, the horizontal axis is the real-axis coordinate along $x$, the longitudinal axis is energy and the brightness stands for the relative strength of DOS. There are two nearly degenerate states near the Fermi energy. Two lower panels are the real space LDOS distribution of the two junction states.}\label{fig2}
\end{figure}

To check the topological index of the junction state, we calculate the electronic structure of $L_{1}$-$L_{2}$-$N_{1}$-$N_{2}$-GNRHs using the tight-binding simulation with fixed $L_{1}=L_{2}=8$ to prevent overlap between the neighboring junction states. In the $8$-$8$-$9$-$17$-GNRH with NM configuration, the topological index of the junction state is $\Delta\mathbb{Z}=2$. The bandstructure exhibits two branches (two pairs of valence/conduction bands) of near-Fermi-Level band (Fig.~\ref{fig2} (f)). The chain-resolved local density of states (LDOS) show that they are mainly localized at the junctions where two AGNRs connected (Fig.~\ref{fig2} (a)), manifesting their junction state nature. Moreover, the real-space-resolved LDOS distribution reveals that the two junction states locate on $A$ sublattice on left junction, while on $B$ sublattice on right junction where $\Delta\mathbb{Z}=-2$ due to exchange of $N_{1}$ and $N_{2}$ ((Fig.~\ref{fig2} (b) and (c)). Such chirality further confirms the topological nature of those junction states (More in SM). In comparison, the bulk state almost equally distributes along the ribbon. Similar junction states with $\Delta \mathbb{Z}=2$ can also be observed in the corresponding AFMB configuration (right panels in Fig.~\ref{fig2}). Therefore, the emergent magnetism does not change the numbers and chirality of topological junction states. However, unlike the NM configuration, the spin-resolved junction states mainly locate on the one of edges along the ribbon due to strong edge magnetization in magnetic configurations, which will substantially change the topological phase diagram of GNRH as shown below.  Additionally, the bandgap in AFMB configuration ($\sim 0.5$ eV) is much larger than that in NM configuration ($\sim 0$ eV), supporting the robustness of the topological junction states (More in SM).

The existence and the number of the topological junction states can also be understood in a simple picture in term of gap closing and reopening. The bulk states of ($3N+2$)-AGNR are gapless, while the ($3N+1$)-AGNR and $3N$-AGNR are gapped~\cite{Nakada-PRB1996,Son-PRL2006}, the number of junction states can be simplified by counting the number of gap closure from $N_1$ to $N_2$. For example, when $N_1=9, N_2=17$, the gap closes twice at $N=11,14$, resulting in two junction states. When $N_1=9, N_2=13$, the gap closes only once at $N=11$, yielding one topological junction state. The result is the same as that obtained by using $\mathbb{Z}$ invariants.

\begin{figure}[!htb]
  \centering
  \includegraphics[width=\linewidth]{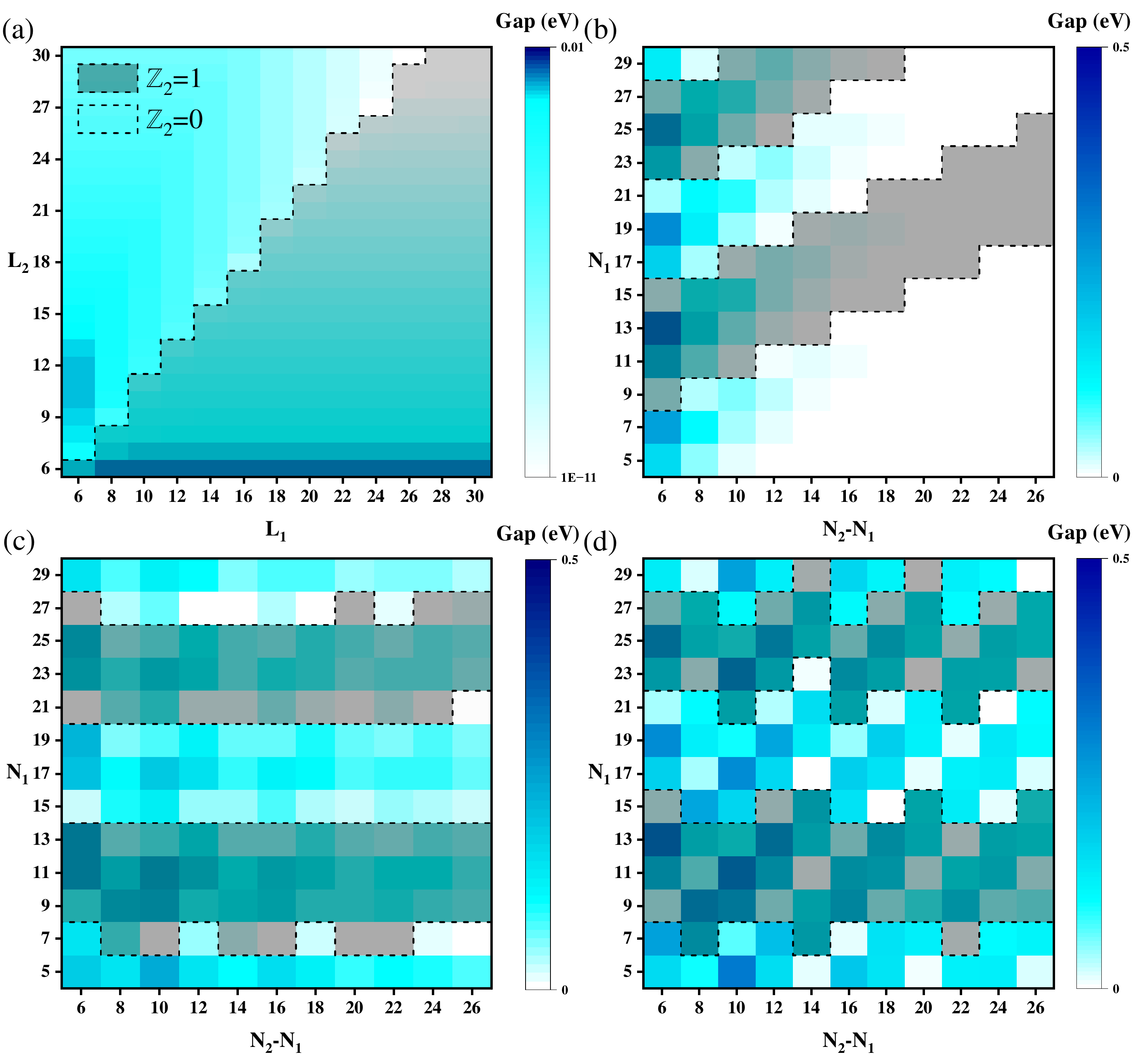}
  \caption{Topological phase diagrams of GNRH. (a) and (b) for the NM configuration, (c) for AFMB configuration, and (d) for AFMC configuration. (a) Topological phase diagram with respect to $L_1$ and $L_2$ for fixed $N_{1}=9, N_{2}=17$. (b), (c) and (d) Topological phase diagram with respect to $N_1$ and $N_2 - N_1$ with fixed $L_{1}=2, L_{2}=3$. The gray, and blank area, denote the topologically non-trivial ($\mathbb{Z}_{2}=1$), and trivial phase ($\mathbb{Z}_{2}=0$), respectively. The blue colored scale indicates the magnitude of bandgap. The topological phase diagram for magnetic configurations is insensitive to $L_{1/2}$ in (c) and (d). The AFMA configuration is always a normal insulator.}\label{fig3}
\end{figure}

\emph{Bulk topology and topological phase diagram}---We now study the bulk topology of the designed GNRHs. We calculate the $\mathbb{Z}_2$ invariants using both Zak phase and parity eigenvalue (SM for details) for the NM, AFMB, and AFMC configurations, directly from mean-field theory due to their preservation of inversion or/and mirror symmetry. Since the AFMA configuration preserves neither the mirror nor the inversion symmetry, the wavefunctions have no parity eigenvalues, and the corresponding Zak phase is not well-defined. Fig.~\ref{fig3} summarizes the topological phase diagrams.

In NM configuration, the topological phase is highly geometrically dependent. The topologically nontrivial phase in $L_{1}$-$L_{2}$ space, forming a triangular shape, emerges when $L_{1}>L_{2}$ for fixed $N_{1}=9, N_{2}=17$ (Fig.~\ref{fig3} (a)). This length-dependent phase diagram well agree with the inline-type heterojunctions with $N_{1}=7$ and $N_{2}=9$ reported before~\cite{Groning-Nature2018}. Moreover, we find that the bulk topology is also sensitive to the width $N_{1}$ and $N_{2}$ of two AGNR segments. A diagonal-stripe-like topologically trivial and nontrivial alternation, with approximately $N=12$-period, is observed in $N_{1}-N_{2}$ space (Fig.~\ref{fig3} (b)). Such $12$-period perhaps relates to the two ZGNR segments unit cells ($N_{1}$ and $N_{2}$), each has the $\mathbb{Z}_{2}=\frac{1+(-1)^{\lfloor{\frac{N}{3}}\rfloor+\lfloor{\frac{N+1}{2}}\rfloor}}{2}$ topological classification of AGNRs with closed zigzag boundary as revealed by Cao \emph{et al.}~\cite{Cao-PRL2017}, where a $N=6$-alternation is expected for each due to the two fractions in floor function in the formula. The geometric sensitivity provides a way to engineer the bulk topology of GNRHs. We would like to point out that the bandgap of $\mathbb{Z}_{2}$ phase in NM is indeed narrow, especially for larger width, rendering the topological phases may be unstable at higher temperature or with disorders.

Surprisingly, the topological phase in magnetic configurations depends solely on the narrow AGNR width $N_{1}$ (Fig.~\ref{fig3} (c) and (d)). Similar to the NM case, the phase alternation with $12$-period (but only $N_{1}$) are also observed. This highlights a shared topological origin between NM and magnetic configurations, with magnetism playing a crucial role in shaping the topology. We notice that there are some irregulars on the phase boundary. It maybe relate to uncertainty in the bandgap closing and reopening due to the discrete changes of $N_{1}$. The magnetic phase diagrams resemble that of AGNRs with closed zigzag boundaries, where the phase boundary in the latter is $6n+11$ ($n$ is an integer) according to the $\mathbb{Z}_{2}$ index $\mathbb{Z}_2=\frac{1+(-1)^{\lfloor{\frac{N}{3}}\rfloor+\lfloor{\frac{N+1}{2}}\rfloor}}{2}$~\cite{Cao-PRL2017}. Here, the phase boundary shifts to $6n+7$. This suggests the emergence of junction states forms some sorts of equivalent AGNRs but with additional ``phase shift''. In addition, the bandgap in magnetic configurations remains wide even for the large sizes. Our results not only provide a way to realize the robust topology, but also offer an accessible way to control the topology in GNRHs by means of magnetism.

\begin{figure}[!htb]
  \centering
  \includegraphics[width=\linewidth]{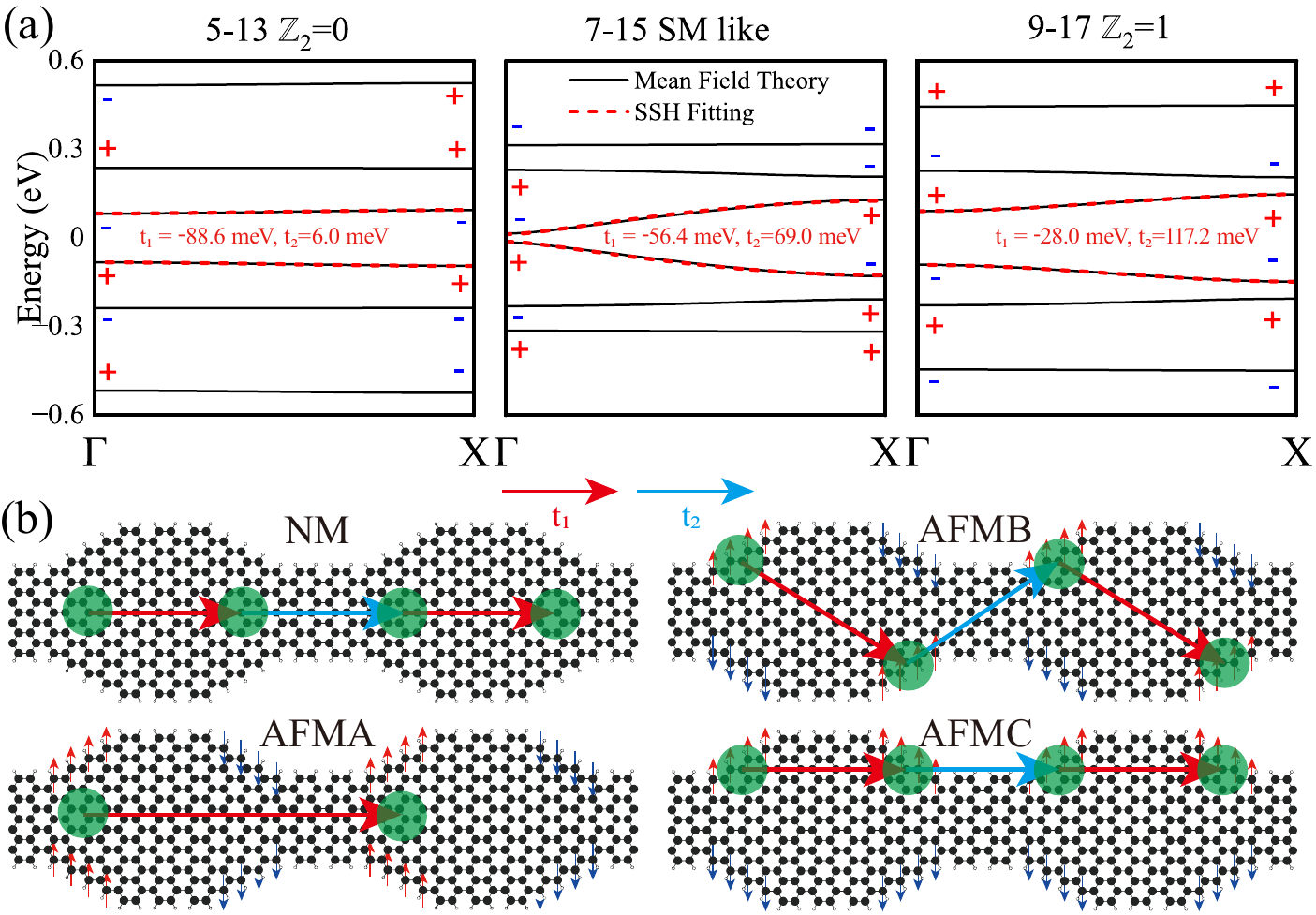}
  \caption{(a) From left to right, the band structures evolution with fixed $L_{1}=2$ and $L_{2}=3$ under the mean field approximation. The parity of the bands at high symmetry points is marked with ``$+$'' (even) and ``$-$'' (odd). The red dash line shows the SSH fitting results. The fitted parameters are shown inset. (b) The positions of the WCs and the schematic hopping process between WCs for the four possible configurations. Red, and blue arrows represent the intra- ($t_{1}$) and inter-dimer ($t_{2}$) hopping, respectively. For magnetic configurations, only the spin-up hopping channel is shown.}\label{fig4}
\end{figure}

The $\mathbb{Z}_{2}$ topological phase transition can be also identified by the band inversion~\cite{Qi-RMP2011}. We show the typical process of the near-Fermi-level bandstructure evolution on $N_{1}$ with fixed $N_{2}-N_{1}=8$ in AFMB configuration using the $\pi$-orbital Hubbard model under the mean-field approximation (Fig.~\ref{fig4}(a)). The process of bandgap closing and reopening is accompanied by the switching of parity eigenvalues of the valence and conduction bands at high-symmetry points $\Gamma$ and $X$. Such band inversion naturally yields the $\mathbb{Z}_{2}$ invariant changing between $0$ and $1$. 

\emph{Effective SSH mechanism}---As shown above, topological junction state always exists near the junction between two AGNRs in GNRHs. The topological quantum phases in GNRH in the NM state can be understood using the effective SSH model~\cite{Rizzo-Nature2018,Groning-Nature2018}, where the effective coupling between junction states is represented by hopping amplitudes $t_1$ (across the wider $N_2$-AGNR segment) and $t_2$ (across the narrow $N_1$-AGNR segment). Since each unit cell contains two inequivalent junction states, the dispersion arising from the GNHR system can be modeled as $E_{\pm}(k)=\pm\sqrt{t_{1}^{2}+t^{2}_{2}+2t_{1}t_{2}\cos(k)}$. The frontier bands shown in Fig.~\ref{fig4}(a) are fitted using the effective hopping parameters $t_{1}$ and $t_{2}$ shown in the inset. For $2$-$3$-$5$-$13$-GNRH, $\vert t_{1}/t_{2}\vert > 1$, verifying a topologically trivial state. For $2$-$3$-$7$-$15$-GNRH, $\vert t_{1}/t_{2}\vert \sim 1 $, indicating on the topological phase transition boundary. For the $2$-$3$-$9$-$17$-GNRH, $\vert t_{1}/t_{2}\vert < 1$ supporting a topologically nontrivial state. The topological phase transition analysed by the effective SSH mechanism well agrees with the above band inversion picture. We, therefore, show that the topology in GNRH with magnetic configurations can also be well described by the effective SSH model, where the junction states serve as dimers, and $t_{1}$ and $t_{2}$ can be viewed as intra- and inter-dimer hopping, respectively. Here, the effective hopping parameters have opposite signs, which is consistent with the direct band gap at the $\Gamma$ point and similar to previously reported parameters with NM configuration~\cite{Rizzo-Nature2018}.

To understand the difference between the topological phase diagram of NM configuration and that of magnetic configurations, we resort to the WCs. In Fig.~\ref{fig4}(b), we calculate the maximally localized Wannier functions for the pair of conduction and valence bands closest to the Fermi level directly from the first-principle simulations. For the NM configuration, the WCs locate at the junctions along the center of ribbon. In contrast, the WCs (spin-resolved) locates at junction along the edge of ribbon. The effective hopping parameters between the WCs are shown in Tab.~S2 (SM), the relative magnitude of $t_{1}$ and $t_{2}$ are well consistent with the topological phase diagram (Fig.~\ref{fig3}), confirming the SSH mechanism of the topology in GNRHs.

Due to the absence of spin-flip hopping mechanism in the present heterojunctions, each configuration shows distinct hopping process (colored arrows in Fig.~\ref{fig4}(b)). For AFMA configuration, no SSH mechanism exist due to the absence of the inter-dimer hopping process, which naturally explains the absence of SPT phase. In NM configuration, both the intra- and inter-dimer hopping process are sensitive to the width and length of AGNRs. In contrast, the effective hopping parameters in magnetic configurations are sensitive to the width, but less sensitive to the lengths of AGNRs, especially their relative magnitudes are almost insensitive to the length of AGNRs. These features are qualitatively consistent with the main features of the obtained topological phase diagrams.

\begin{figure}[!htb]\centering\includegraphics[width=\linewidth]{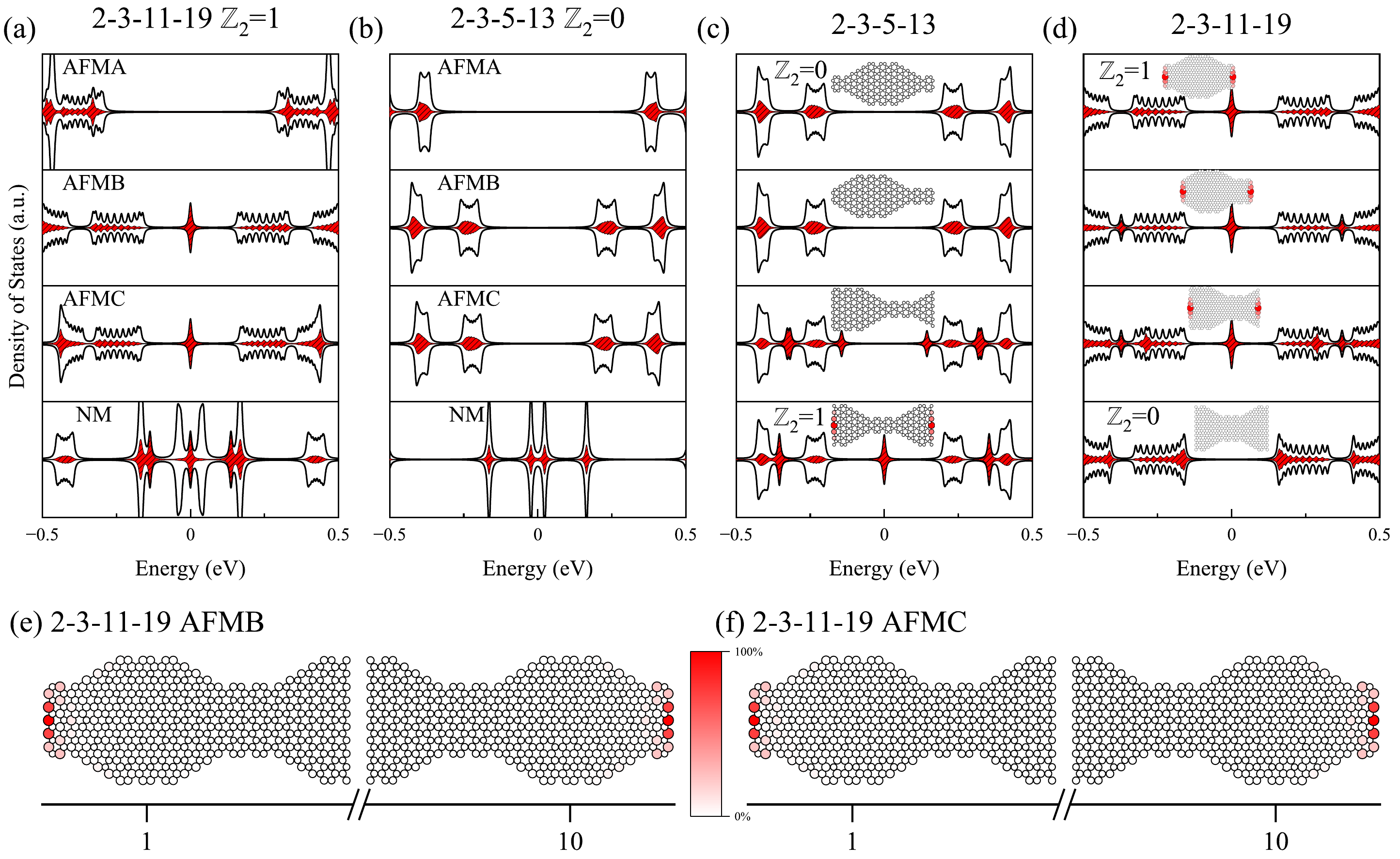}\caption{Topological end states. (a)DOS of $2$-$3$-$11$-$19$-GNRH obtained by the $\pi$-orbital Hubbard model simulations. The grey shaded area represents the DOS contribution of end cell. A finite $2$-$3$-$11$-$19$-GNRH with ten unitcells is adopted. Here, $\mathbb{Z}_{2}=1$ for AFMB, AFMC, and NM configurations, while AFMA is not a $\mathbb{Z}_{2}$ topological insulator. The broaden factor is set to 5 meV. (b) Same as (a), but for $2$-$3$-$5$-$13$-GNRH. (c)DOS of $2$-$3$-$11$-$19$-GNRH with ten unitcells, the corresponding unitcell is changed and shown inset. (d)Same as (c), but for $2$-$3$-$5$-$13$-GNRH. (e) The real space LDOS of the zero mode end states of $2$-$3$-$11$-$19$-GNRH in AFMB phase. (f) Same as (e), but for $2$-$3$-$11$-$19$-GNRH in AFMC phase. }\label{fig5}\end{figure}

\emph{Topological end states}---A topologically nontrivial phase hosts unique edge states due to the bulk-edge correspondence. In our topological GNRH, we expect the emergence of the topological end states at terminations of finite GNRH. To verify, we calculate the total density of states (TDOS), as well as the density of states (DOS) contributed from the end cells of a finite GNRH containing ten unitcells. Here the DOS of the end cells is defined as the projected DOS of the left end unitcell and the right end unitcell. We show the typical results with the topological index $\mathbb{Z}_{2}=1$ in Fig.~\ref{fig5}(a). In the NM configuration, we observe a zero-mode contributed by the end unitcell, but it is mixed with the bulk state. This again indicates the instability of topological phase in NM configuration, agreeing with the above mentioned narrow bandgap in energy spectrum. In contrast, a well-defined zero-mode completely contributed by the end unitcell emerges in AFMB and AFMC configurations. The zero-mode is well separated from the bulk DOS, supporting a stable topologically nontrivial phase. By the way, no zero-mode can be found in AFMA configuration at any parameters, consisting with the absence of SPT phase. We project the zero-mode in AFMB and AFMC configurations in real space(Fig.~\ref{fig5}(e) and (f)), and identify it well locates on the two terminations of GNRH. For comparison, Fig.\ref{fig5}(b) shows the DOS for all four configurations with $\mathbb{Z}_2 = 0$, where no zero-mode or in-gap state is observed, reflecting normal insulator behavior. Those features confirm the emergence of robust $1$D topological phase in magnetic GNRHs, as well as the topological phase transition.

In principle, the edge modes is robust against the boundary geometry in two- or three-dimensional topological insulators. However, the zero-dimensional end states are sensitive to the termination geometries~\cite{Su-PRB1980}. We show such sensitivity in Fig.~\ref{fig5}(c) and (d). For example, for $2$-$3$-$11$-$19$-GNRH with AFMB configuration, no zero-mode, \emph{i.e.,} no topological end state is observed for the original geometry, consisting with the topological index $\mathbb{Z}_{2}=0$. When the unitcell changes gradually with the junction state shifting left, the in-gap state develops. In particular, after half-unitcell shift, a well-defined zero-mode located on the termination is observed, manifesting the emergence of topological end state. Similar alternation is also observed in $2$-$3$-$11$-$19$-GNRH with AFMB configuration ($\mathbb{Z}_{2}=1$), the topological end state in the case of original geometry vanishes after half-unitcell shift. Such alternation well support the SSH mechanism in bulk topology of GNRH, in which the effective $t_{1}$ and $t_{2}$ exchange with each other. The sensitivity to termination geometry also provides tunable way to engineer the topological end state in $1$D GNRHs.

\emph{Discussions}---In conclusion, we study the role of magnetism on the topology in $1$D graphene nanoribbon heterojunctions based on the first-principle and model simulations. We reveal that the zigzag-edge induced antiferromagnetism does not change the number and chirality of topological junction states in non-magnetic state if the space symmetry is preserved. However, it dramatically changes the location of spin-resolved WCs, which forms new scheme of SSH mechanism. The unique magnetism-tuned topological phase diagram only depends on the width of the narrow AGNR. Moreover, the strong edge magnetism significantly improve the robustness of topological phase. We also show that the topological end states can be readily manipulated by the termination of ribbons. Our results, therefore, provide an accessible way to realize robust topology and engineer the topology in $1$D graphene nanoribbon heterojunctions.

Experimentally, the proposed GNRHs can be atomically precisely fabricated by the on-surface synthesis or bottom-up technique\cite{Cai-Nature2010,Ruffieux-Nature2016}. Therefore, these potential phases and topological end states can be detected by the scan tunneling spectroscopy measurement. The desired magnetic order can be obtained by applying magnetic field. One can also tune the topological properties by applying electric field or strain field\cite{Zhao-PRL2021,Huang-PRB2024}.

\emph{Acknowledgments}---This work is financially supported by the National Natural Science Foundation of China (Grant Nos. 12374137 (YZ), 12204404 (ALH), and 51972217 (SJ)). YZ also acknowledges the Natural Science Foundation of Jiangsu Province (Grant No. BK20231397). SJ also thanks Suzhou Basic Research Project (SJC2023003 (SJ)), and Wenzheng Scholarship of Suzhou City University.

\bibliography{sn-bibliography}
\end{document}